# The complex structure of Abell 2345: a galaxy cluster with non–symmetric radio relics

W. Boschin[1], R. Barrena[2,3,4], and M. Girardi[4,5]

[1] Fundación Galileo Galilei - INAF, Rambla José Ana Fernández Perez 7, E-38712 Breña Baja (La Palma), Canary Islands, Spain

[2] Instituto de Astrofísica de Canarias, C/Vía Láctea s/n, E-38205 La Laguna (Tenerife), Canary Islands, Spain

[3] Departamento de Astrofísica, Universidad de La Laguna, Av. del Astrofísico Francisco Sánchez s/n, E-38205 La Laguna (Tenerife), Canary Islands, Spain

[4] Dipartimento di Fisica dell'Università degli Studi di Trieste - Sezione di Astronomia, via Tiepolo 11, I-34143 Trieste, Italy

[5] INAF - Osservatorio Astronomico di Trieste, via Tiepolo 11, I-34143 Trieste, Italy

**ABSTRACT**

*Context.* The connection of cluster mergers with the presence of extended, diffuse radio sources in galaxy clusters is still debated.

*Aims.* We aim to obtain new insights into the internal dynamics of the cluster Abell 2345. This cluster exhibits two non-symmetric radio relics well studied through recent, deep radio data.

*Methods.* Our analysis is based on redshift data for 125 galaxies acquired at the Telescopio Nazionale Galileo and on new photometric data acquired at the Isaac Newton Telescope. We also use ROSAT/HRI archival X–ray data. We combine galaxy velocities and positions to select 98 cluster galaxies and analyze the internal dynamics of the cluster.

*Results.* We estimate a mean redshift $\langle z \rangle = 0.1789$ and a line–of–sight (LOS) velocity dispersion $\sigma_V \sim 1070$ km s$^{-1}$. The two–dimensional galaxy distribution reveals the presence of three significant peaks within a region of $\sim 1\ h_{70}^{-1}$ Mpc (the E, NW, and SW peaks). The spectroscopic catalog confirms the presence of these three clumps. The SW and NW clumps have similar mean velocities, while the E clump has a larger mean velocity ($\Delta V_{\rm rf} \sim 800$ km s$^{-1}$); this structure causes the presence of the two peaks we find in the cluster velocity distribution. The difficulty in separating the galaxy clumps leads to a very uncertain mass estimate $M \sim 2 \times 10^{15}\ h_{70}^{-1}\ M_\odot$. Moreover, the E clump well coincides with the main mass peak as recovered from the weak gravitational lensing analysis and is off–set to the east from the BCG by $\sim 1.3'$. The ROSAT X–ray data also show a very complex structure, mainly elongated in the E–W direction, with two (likely three) peaks in the surface brightness distribution, which, however, are off–set from the position of the peaks in the galaxy density. The observed phenomenology agrees with the hypothesis that we are looking at a complex cluster merger occurring along two directions: a major merger along the ~E–W direction (having a component along the LOS) and a minor merger in the western cluster regions along the ~ N–S direction, roughly parallel to the plane of the sky. The eastern radio relic is elongated in the direction perpendicular to that of the major merger, while the peculiar, western radio relic is elongated in the direction perpendicular to the bisecting of the two merger directions.

*Conclusions.* Our scenario for the internal dynamics of Abell 2345 strongly supports the use of the "outgoing merger shocks" model to explain the two radio relics, suggesting a consistent justification for their asymmetry.

# 1. Introduction

Diffuse radio emission in galaxy systems is a rare phenomenon presently known in only a few tens of (mostly) rich clusters (e.g. Giovannini et al. 1999; see also Giovannini & Feretti 2002; Feretti 2005a; Giovannini et al. 2009). These radio sources (named halos and relics) are very interesting for their possible connection with cluster mergers (see Feretti et al. 2002b for a review). They show a typical synchrotron spectrum, a clear sign of the existence of large–scale cluster magnetic fields and of widespread relativistic particles. Cluster mergers have been suggested to provide the large amount of energy necessary for electron reacceleration up to relativistic energies and for magnetic field amplification (Feretti 1999; Feretti 2002a; Sarazin 2002). Radio relics ("radio gischts" as referred by Kempner et al. 2003), which are polarized and elongated radio sources located in the cluster peripheral regions, seem to be directly associated with merger shocks (e.g., Ensslin et al. 1998; Roettiger et al. 1999; Ensslin & Gopal–Krishna 2001; Hoeft et al. 2004). Radio halos, unpolarized sources which permeate the cluster volume similarly to the X–ray emitting gas, are more likely to be associated with the turbulence following a cluster merger (Cassano & Brunetti 2005). However, only recently the number of diffuse radio sources discovered in clusters has grown enough to allow their study on the basis of a sufficient statistics and to attempt a classification (e.g. Kempner et al. 2003; Ferrari et al. 2008).

There is growing evidence of the connection between diffuse radio emission and cluster mergers based on X–ray data (see Buote 2002; Feretti 2006 and 2008 and refs. therein). Optical data are a powerful way to investigate the presence and the dynamics of cluster mergers (e.g., Girardi & Biviano 2002), too. The spatial and kinematical analysis of member galaxies allow us to detect and measure the amount of substructure, to identify and analyze possible pre–merging clumps or merger remnants. This optical information is really complementary to X–ray information since galaxies and intra–cluster medium react on different time scales during a merger (see, e.g., numerical simulations by Roettiger et al. 1997). In this context we are conducting an intensive observational and data analysis program to study the internal dynamics of clusters showing extended radio emission by using member galaxies (DARC, Dynamical Analysis of Radio Clusters; Girardi et al. 2007; http://adlibitum.oat.ts.astro.it/girardi/darc).

Among the clusters hosting diffuse radio emission, only a very small fraction is characterized by double radio relics: Abell 1240 (Giovannini et al. 1999; Bonafede et al. 2009, hereafter B09);

---

*Send offprint requests to*: W. Boschin, e-mail: `boschin@tng.iac.es`

Abell 2345 (Giovannini et al. 1999; B09); Abell 3667 (Röttgering et al. 1997); Abell 3376 (Bagchi et al. 2006); RXCJ 1314.4–2515 (Feretti et al. 2005b; Venturi et al. 2007). According to van Weeren et al. (2009) cluster ZwCl 2341.1+0000 could host two radio relics, too. However, recent data by Giovannini et al. (2010) suggest a more complex picture, in which the presumed relics (separated in the high–resolution image of van Weeren et al.) are connected to form a large–scale filament of diffuse radio emission. The relics of Abell 1240 and 3667 were explained by the "outgoing merger shocks" model (Roettiger et al. 1999; B09). Observations of Abell 3376 agree with both the "outgoing merger shocks" and the "accretion shock" models (Bagchi et al. 2006).

In the case of Abell 2345 (hereafter A2345), the observations are more difficult to be reconciled with the theoretical scenarios. In fact, the two relics, separated by $\sim 2\ h_{70}^{-1}$ Mpc, are not perfectly symmetric with respect to the cluster center. Moreover, while the eastern relic (A2345–2) is well explained in the scenario of the "outgoing merger shocks" model, the western relic (A2345–1) is quite peculiar (B09). In particular, the spectral index of A2345–1 steepens in the southern and external parts while diffusive shock acceleration models predict a steepening of the radio spectrum towards the cluster center as a consequence of the electron energy losses after shock acceleration (e.g., Ensslin et al. 1998). B09 suggested that the radio properties of A2345–1 could be affected by a possible ongoing merger with another group lying in the external cluster regions, at the NW with respect to the center of A2345 and visible in ROSAT/PSPC X–ray data (the "X1" group).

During our observational program we have conducted an intensive study of the cluster A2345 performing spectroscopic and photometric observations with the Telescopio Nazionale Galileo (TNG) and the Isaac Newton Telescope (INT), respectively.

A2345 is a very rich, X–ray luminous, Abell cluster: Abell richness class = 2 (Abell et al. 1989); $L_X$(0.1–2.4 keV)=$4.28 \times 10^{44}\ h_{70}^{-2}$ erg s$^{-1}$ recovered from ROSAT data (Böhringer et al. 2004). Optically, the cluster is classified as Bautz–Morgan class III (Abell et al. 1989), while it is classified as cD cluster according to Rood–Sastry scheme (Struble & Rood 1987).

Weak gravitational lensing analysis has been performed by Dahle et al. (2002; hereafter D02) and by Cypriano et al. (2004) for the region close to the brightest cluster galaxy (hereafter BCG). D02 found that the highest peak in the mass map is off–set to the east from the central BCG by $\sim 1.5'$ and noticed that ROSAT/HRI archival data show a large amount of substructure.

Few galaxies having redshift are reported in the field of A2345 (see NED) and the value usually quoted in the literature for the cluster redshift comes from the spectrum of 2124–120, a radio galaxy at $z = 0.176$ (Owen et al. 1995; Owen & Ledlow 1997).

This paper is organized as follows. We present our new optical data and the cluster catalog in Sect. 2. We present our results about the cluster structure in Sect. 3. We discuss our results and give our conclusions in Sect. 4.

Unless otherwise stated, we give errors at the 68% confidence level (hereafter c.l.). Throughout this paper, we use $H_0 = 70$ km s$^{-1}$ Mpc$^{-1}$ in a flat cosmology with $\Omega_0 = 0.3$ and $\Omega_\Lambda = 0.7$. In the adopted cosmology, 1$'$ corresponds to $\sim 181\ h_{70}^{-1}$ kpc at the cluster redshift.

## 2. New data and galaxy catalog

Multi–object spectroscopic observations of A2345 were carried out at the TNG telescope in August 2008. We used DOLORES/MOS with the LR–B Grism 1, yielding a dispersion of 187 Å/mm. We used the new E2V CCD, a matrix of $2048 \times 2048$ pixels with a pixel size of 13.5 $\mu$m. In total we observed four MOS masks for a total of 147 slits. Total exposure times were of 1 hour for three masks and 1.5 hours for the remaining mask. Wavelength calibration was performed using Mercury–Neon lamps. Reduction of spectroscopic data was carried out with the IRAF [1] package.

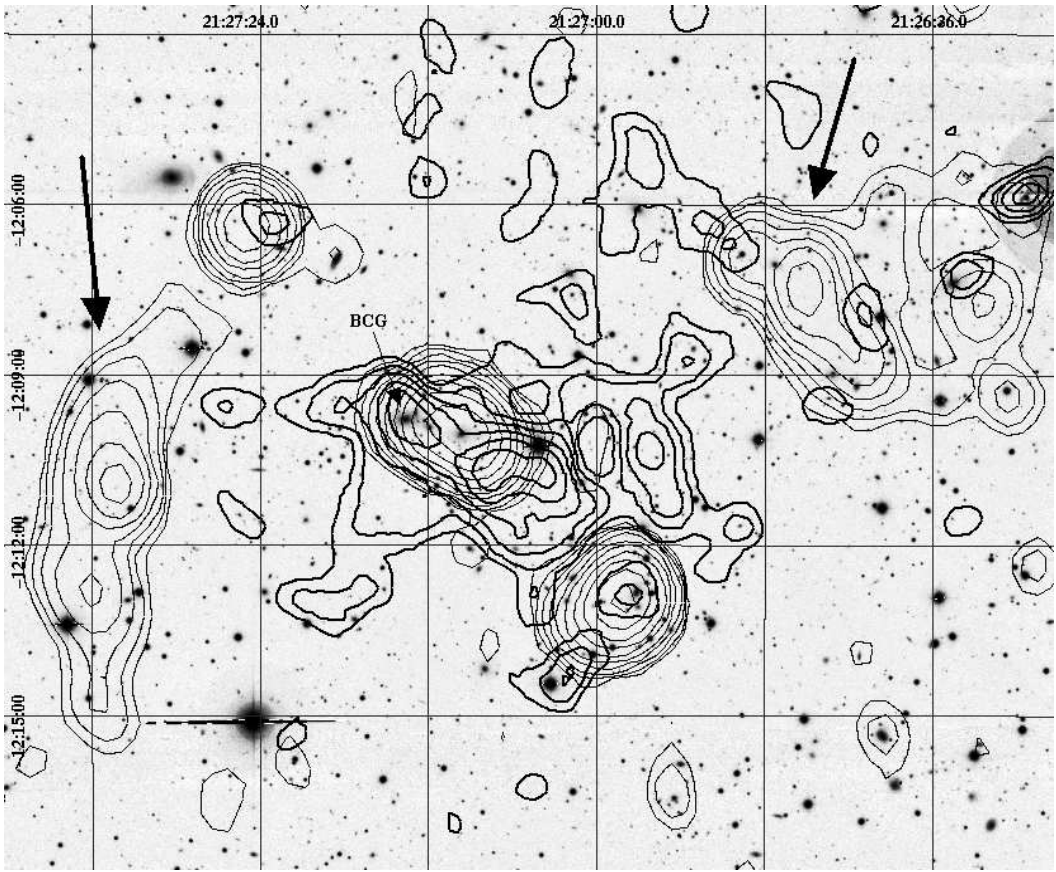

**Fig. 1.** INT $r'$–band image of the cluster A2345 (North at the top and East to the left) with, superimposed, the contour levels of the (smoothed) ROSAT/HRI archival image (pointing ID: US800769H, thick contours) and the contour levels of a VLA radio image at 1.4 GHz (see B09; beam $50''\times38''$, thin contours starting at 0.24 mJy/beam and then spaced by a factor 2). Arrows highlight the contour levels of the two radio relics.

Radial velocities were determined using the cross–correlation technique (Tonry & Davis 1979) implemented in the RVSAO package (developed at the Smithsonian Astrophysical Observatory Telescope Data Center). Each spectrum was correlated against six templates for a variety of galaxy spectral types: E, S0, Sa, Sb, Sc, Ir (Kennicutt 1992). The template producing the highest value of $\mathcal{R}$, i.e., the parameter given by RVSAO and related to the signal–to–noise ratio of the correlation peak, was chosen. Moreover, all spectra and their best correlation functions were examined visually to verify the redshift determination. In two cases (IDs. 34 and 108; see Table 1) we took the EMSAO redshift as a reliable estimate of the redshift. Our spectroscopic survey in the field of A2345 consists of 132 spectra with a median nominal error on $cz$ of 36 km s$^{-1}$. The nominal errors as given by the cross–correlation are known to be smaller than the true errors (e.g., Malumuth et al. 1992; Bardelli et al. 1994; Ellingson & Yee 1994; Quintana et al. 2000; Boschin et al. 2004).

---

[1] IRAF is distributed by the National Optical Astronomy Observatories, which are operated by the Association of Universities for Research in Astronomy, Inc., under cooperative agreement with the National Science Foundation.

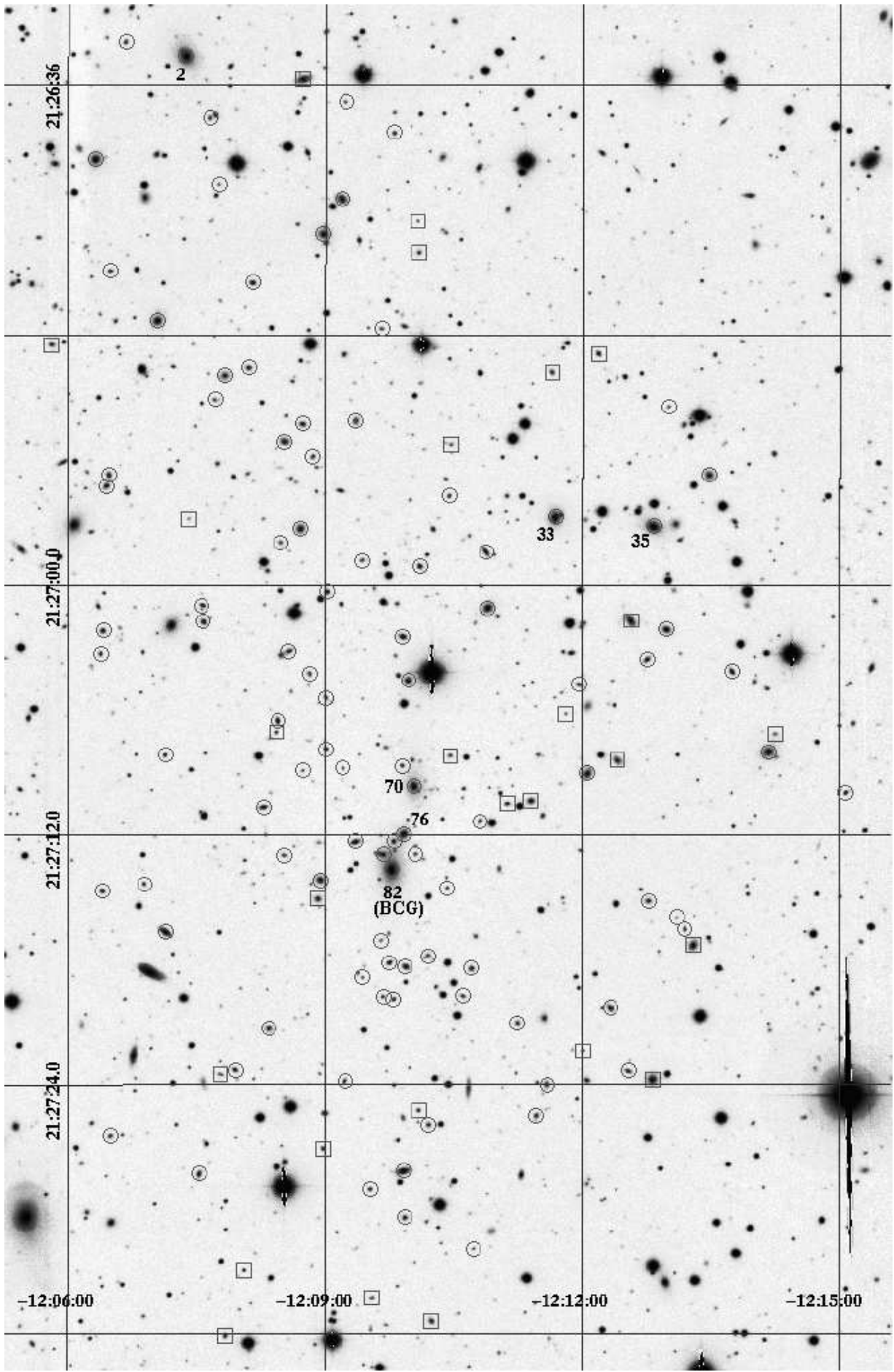

**Fig. 2.** INT $r'$–band image of the cluster A2345 (West at the top and North to the left). Circles and boxes indicate member and non–member galaxies, respectively (see Table 1). Labels indicate the IDs of cluster galaxies cited in the text.

Double redshift determinations for seven galaxies allowed us to estimate real intrinsic errors in data of comparable quality taken with the same instrument (e.g. Barrena et al. 2007a, 2007b). We fit the first determination vs. the second one by using a straight line and considering errors in both coordinates (e.g., Press et al. 1992). The fitted line agrees with the one to one relation, but, when using the nominal cross–correlation errors, the small value of $\chi^2$–probability indicates a poor fit, suggesting the errors are underestimated. Only when nominal errors are multiplied by a $\sim 1.4$ factor the observed scatter can be explained. Therefore, hereafter we assume that true errors are larger than nominal cross–correlation errors by a factor 1.4. For the seven galaxies with two redshift measurements we used the mean of the two redshift determinations and the corresponding errors.

Finally, our spectroscopic survey in the field of A2345 consists of spectra for 125 galaxies with a median error on $cz$ of 50 km s$^{-1}$.

As far as photometry is concerned, our observations were carried out with the Wide Field Camera (WFC), mounted at the prime focus of the 2.5m INT telescope. We observed A2345 with $g'$, $r'$ and $i'$ Sloan–Gunn filters on December 2008 in photometric conditions and seeing of $\sim 1.6''$ ($g'$ and $r'$ images) and $\sim 2.5''$ ($i'$ image).

The WFC consists of a four–CCD mosaic covering a 33′×33′ field of view, with only a 20% marginally vignetted area. We took 9 exposures of 720 s in $g'$ band, 13 of 360 s in $r'$ band and 10 of 600 s in $i'$ band. So, the total exposure times were 6480 s, 4680 s and 6000 s in $g'$, $r'$ and $i'$ band, respectively. By developing a dithering pattern we were able to build a "supersky" frame that was used to correct our images for fringing patterns (Gullixson 1992). In addition, the dithering helped us to clean cosmic rays and avoid gaps between the CCDs in the final images. The complete reduction process (including flat fielding, bias subtraction and bad–column elimination) yielded a final coadded image where the variation of the sky was lower than 0.8% in the whole frame. Another effect associated with the wide field frames is the distortion of the field. In order to match the photometry of several filters, a good astrometric solution is needed to take into account these distortions. Using the *imcoords* IRAF tasks and taking as a reference the USNO B1.0 catalogue, we were able to find an accurate astrometric solution (rms $\sim 0.4''$) across the full frame. The photometric calibration was performed observing the SA92 Landolt standard field, that is also calibrated in the Sloan–Gunn system (Smith et al. 2002).

We finally identified galaxies in our $g'$, $r'$ and $i'$ images and measured their magnitudes with the SExtractor package (Bertin & Arnouts 1996) and AUTOMAG procedure. In a few cases (e.g. close companion galaxies, galaxies close to defects of the CCD) the standard SExtractor photometric procedure failed. In these cases we computed magnitudes by hand. This method consists in assuming a galaxy profile of a typical elliptical galaxy and scaling it to the maximum observed value. The integration of this profile gives us an estimate of the magnitude. This method is similar to PSF photometry, but assumes a E–type galaxy profile, more appropriate in this case.

As a final step, we estimated and corrected the galactic extinction from Burstein & Heiles's (1982) reddening maps. The values of the extinction coefficients are $A_{g'} \sim 0.15$, $A_{r'} \sim 0.12$ and $A_{i'} \sim 0.09$ for the $g'$, $r'$ and $i'$ Sloan–Gunn filters, respectively.

We estimated that our photometric sample is complete down to $g' = 23.3$ (24.0), $r' = 22.7$ (23.3) and $i' = 21.4$ (22.5) for $S/N = 5$ (3) within the observed field.

We measured redshifts for galaxies down to magnitude $r' \sim 21$, but a high level of completeness is reached only for galaxies with magnitude $r' < 20$ (>60% completeness).

Table 1 lists the velocity catalog (see also Fig. 2): identification number of each galaxy, ID (Col. 1); right ascension and declination, $\alpha$ and $\delta$ (J2000, Col. 2); $g'$, $r'$ and $i'$ magnitudes (Cols. 3, 4 and 5); heliocentric radial velocities, v = $cz_\odot$ (Col. 6) with errors, $\Delta$v (Col. 7); member assignment (Col. 8; 1:A2345, 2:background/foreground).

### 2.1. Radio galaxies in the field of A2345

Our spectroscopic catalog lists some radio galaxies observed in the sky region of A2345. Contour levels in Fig. 1 show a pointlike radio source ~40″ W of the BCG. Indeed, a high resolution radio image by B09 (see their Fig. 1) reveals that this source is a blend of two head–tail galaxies (ID. 70 and ID. 76 in our catalog). At ~5′ SW of the BCG, Fig. 1 shows another pointlike source. B09 image reveals that this is the radio source 2124–120 cited above, another head–tail galaxy. Its optical counterpart in our catalog is galaxy ID. 35. Finally, ~5′ NE of the BCG there is a third pointlike source (NVSS J212725-120623). Contrary to galaxies IDs. 70, 76 and 35, which are obvious cluster members (see Sect. 3.1 for details on member selection), this radio galaxy has no counterpart in our optical image and is a likely background object.

## 3. Analysis and Results

### 3.1. Member selection and the question of the cluster center

To select cluster members out of 125 galaxies having redshifts, we follow a two steps procedure. First, we perform the 1D adaptive–kernel method (hereafter DEDICA, Pisani 1993 and 1996; see also Fadda et al. 1996; Girardi et al. 1996). We search for significant peaks in the velocity distribution at >99% c.l.. This procedure detects A2345 as two close peaks at $z \sim 0.1775$ and 0.1806 populated by 55 and 46 galaxies, but largely overlapped since many galaxies (66/101) have a non–negligible probability of belonging to both peaks (see Fig. 3). Out of 24 non members, 12 and 12 galaxies are foreground and background galaxies, respectively.

All the galaxies assigned to the cluster are analyzed in the second step which uses the combination of position and velocity information: the "shifting gapper" method by Fadda et al. (1996). This procedure rejects galaxies that are too far in velocity from the main body of galaxies and within a fixed bin that shifts along the distance from the cluster center. The procedure is iterated until the number of cluster members converges to a stable value. Following Fadda et al. (1996) we use a gap of 1000 km s$^{-1}$ – in the cluster rest–frame – and a bin of 0.6 $h_{70}^{-1}$ Mpc, or large enough to include 15 galaxies.

The choice of the center of A2345 is not obvious. No evident dominant galaxy is present in our catalog, e.g. the BCG (ID. 82) is only 0.4 mag brighter in the $r'$ band than the second brightest galaxy (ID. 2). However, ID. 2 is located in the very north–western cluster region. The eastern central region is someway dominated by the BCG, since the other luminous galaxy (ID. 70) is ~ 1 mag fainter. Probably, that is why this cluster was classified as a cD cluster in the past. Other two luminous galaxies, ID. 35 (the radio galaxy 2124–120 discussed above) and ID. 33, lie in the south–western region. From the X–ray point of view, the center is not obvious, too. Böhringer et al. (2004) list a center very close (~ 40″) to the position of the BCG, but Ebeling et al. (1996) list a ~ 4′ NE center. Indeed, the X–ray map is quite substructured as shown by the the analysis of ROSAT data (B09; D02; see also our following analysis). In view of these difficulties we decide to adopt from the operative and graphical point of view the location of the BCG, which is well defined, as the cluster center [R.A.=$21^h27^m13^s.70$, Dec.=$-12°09'46.1''$ (J2000.0)]. After the "shifting gapper" procedure we obtain a sample of 98 fiducial cluster members (see Fig. 4, top panel).

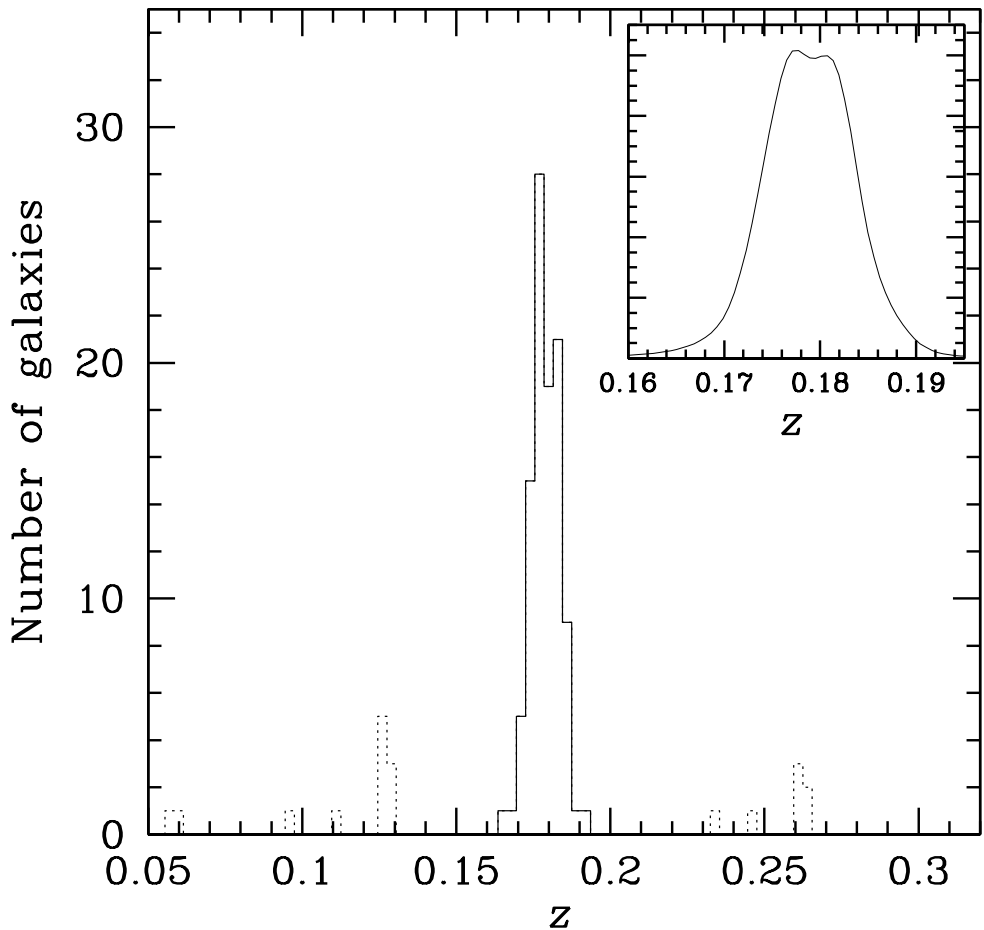

**Fig. 3.** Redshift galaxy distribution. The solid line histogram refers to the 101 galaxies assigned to A2345. The redshift galaxy density, as provided by the adaptive kernel reconstruction method DEDICA, is showed in the insect panel, with an arbitrary normalization.

### 3.2. Global cluster properties

By applying the biweight estimator to the 98 cluster members (Beers et al. 1990, ROSTAT software), we compute a mean cluster redshift of $\langle z \rangle = 0.1789 \pm 0.0004$, i.e. $\langle v \rangle = \langle cz \rangle = (53645 \pm 108)$ km s$^{-1}$. We estimate the LOS velocity dispersion, $\sigma_V$, by using the biweight estimator and applying the cosmological correction and the standard correction for velocity errors (Danese et al. 1980). We obtain $\sigma_V = 1069^{+75}_{-62}$ km s$^{-1}$, where errors are estimated through 1000 bootstrap simulations performed with the ROSTAT routine by Beers et al. (1990), which uses bias–corrected percentile intervals.

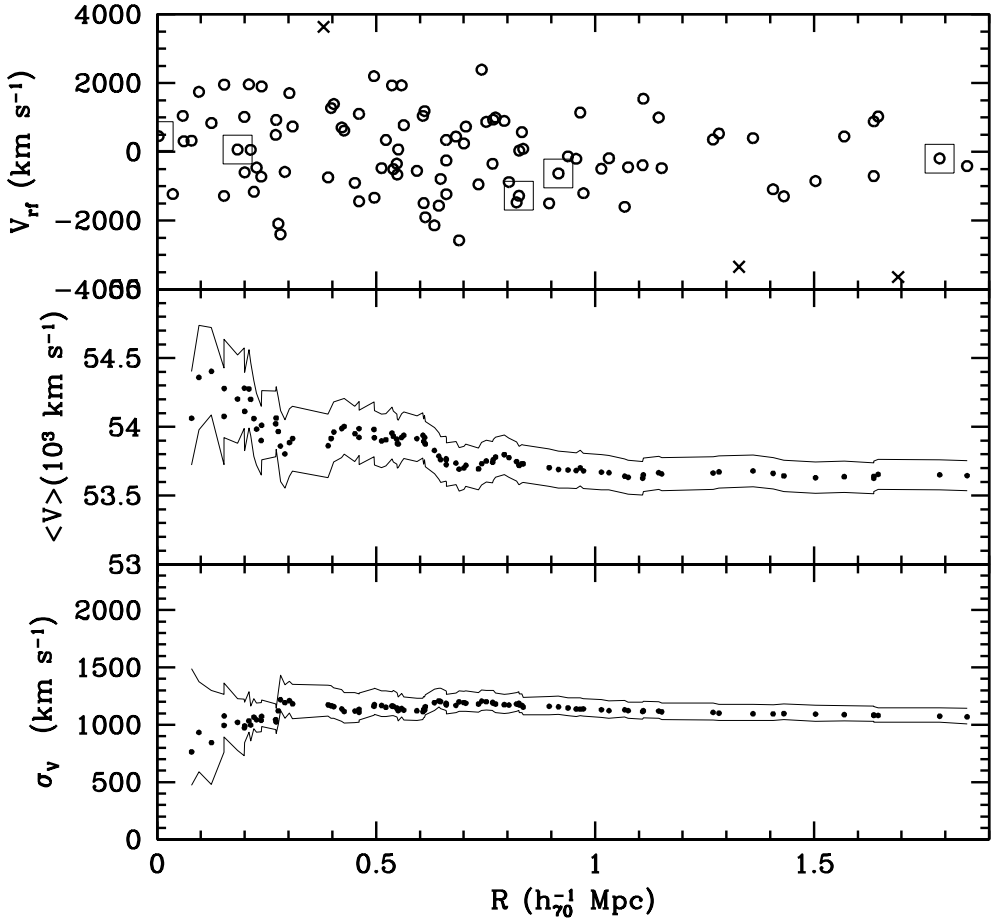

**Fig. 4.** *Top panel:* rest–frame velocity vs. projected clustercentric distance for the 101 galaxies in the two peaks forming the cluster (see Fig. 3). Crosses indicate galaxies rejected as interlopers by our "shifting gapper" procedure. Large squares indicate the five brightest galaxies. *Middle and bottom panels:* integral profiles of mean velocity and LOS velocity dispersion, respectively. The mean and dispersion at a given (projected) radius from the cluster–center are estimated by considering all galaxies within that radius – the first values computed on the five galaxies closest to the center. The error bands at the 68% c.l. are also shown.

To evaluate the robustness of the $\sigma_V$ estimate we analyze the velocity dispersion profile (Fig. 4, bottom panel). The integral profile rises out to $\sim 0.3\ h_{70}^{-1}$ Mpc and then flattens suggesting that a robust value of $\sigma_V$ is asymptotically reached in the external cluster regions, as found for most nearby clusters (e.g., Fadda et al. 1996; Girardi et al. 1996).

### 3.3. Velocity distribution

We analyze the velocity distribution to look for possible deviations from Gaussianity that might provide important signatures of complex dynamics. For the following tests the null hypothesis is that the velocity distribution is a single Gaussian.

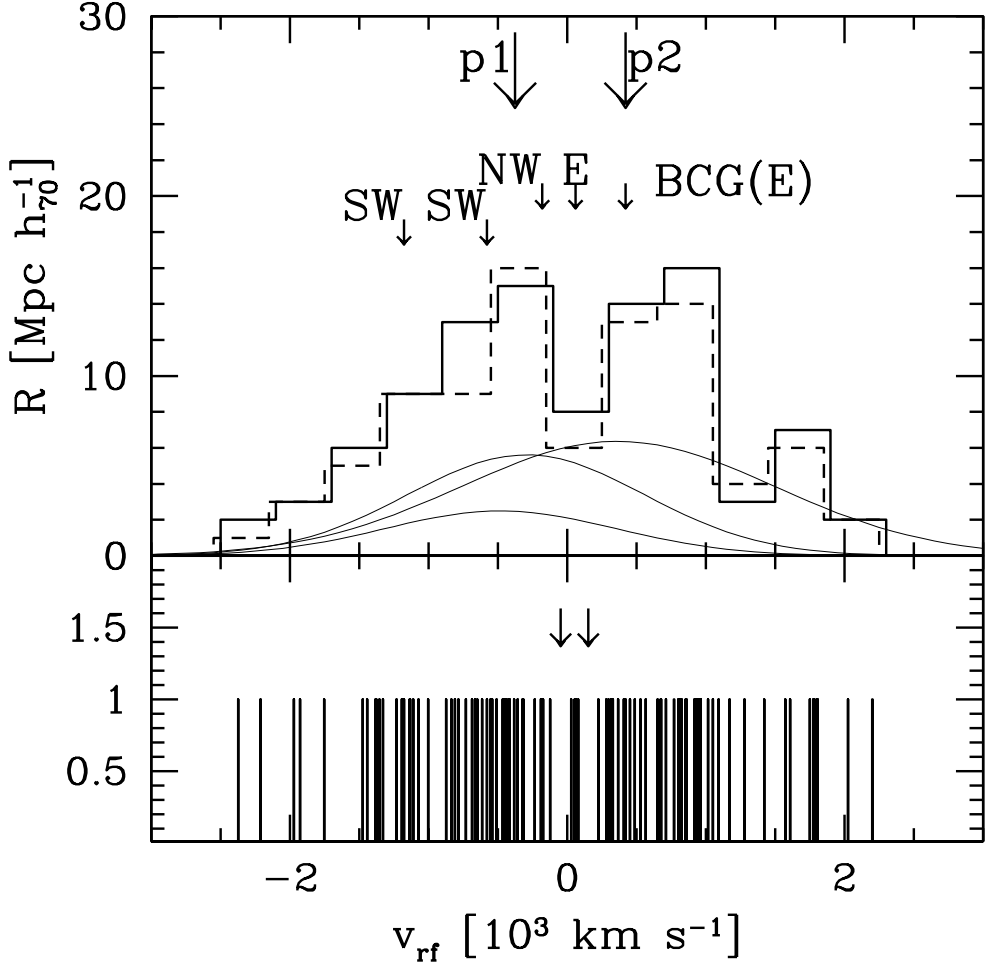

**Fig. 5.** The 98 galaxies assigned to the cluster. *Upper panel*: Velocity distribution for all galaxies and for galaxies with $r' < 20$ (solid and dashed histograms, respectively). The big arrows indicate the position of the two peaks found by the 1D DEDICA algorithm. The Gaussians corresponding to the three galaxies clumps detected in the $r' < 20$ sample trough the 3D DEDICA algorithm are shown (NW, SW and E clump from left to right, respectively). Notice that the separation of these three groups is much more evident in the 2D galaxy distribution (see Fig. 9 in the following). The small arrows show the velocities of the five brightest galaxies indicating the correspondent parent clump. *Lower panel*: Stripe density plot where the arrows indicate the positions of the significant gaps.

We estimate three shape estimators, i.e. the kurtosis, the skewness, and the scaled tail index (see, e.g., Bird & Beers 1993). We find no evidence that the velocity distribution departs from Gaussianity.

Then we investigate the presence of gaps in the velocity distribution. We follow the weighted gap analysis presented by Beers et al. (1991; 1992; ROSTAT software). We look for normalized gaps larger than 2.25 since in random draws of a Gaussian distribution they arise at most in about 3% of the cases, independent of the sample size (Wainer and Schacht 1978). We detect two significant gaps (at the 99.4% c.l.) which divide the cluster in three groups of 48, 5 and 45 galaxies from low to high velocities (see Fig. 5, lower panel).

We also use here the Kaye's mixture model (KMM) test to find a possible group partition of the velocity distribution (as implemented by Ashman et al. 1994). The KMM algorithm fits a user–specified number of Gaussian distributions to a dataset and assesses the improvement of that fit over a single Gaussian. In addition, it provides the maximum–likelihood estimate of the unknown n–mode Gaussians and an assignment of objects into groups. This algorithm is usually used to analyze the velocity distribution where theoretical and/or empirical arguments indicate that the Gaussian model is reasonable. Here the one–dimensional KMM test fails in confirming both the two groups suggested by the DEDICA analysis and the three groups detected by the weighted gap method.

*3.4. 2D cluster structure*

To analyze the two dimensional galaxy distribution we first use our photometric data sample which covers a larger spatial region.

In our photometric catalog we select likely members on the base of both ($r'$–$i'$) and ($g'$–$r'$) colours. Goto et al. (2002) showed that there is a small tilt in the color–magnitude relations ($r'$–$i'$) vs. $r'$ and ($g'$–$r'$) vs. $r'$ and that the scatter in the latter relation is roughly the double than the scatter in the first one. Out of our photometric catalog we consider as likely cluster members those objects with a SExtractor stellar index ≤ 0.9 having ($r'$–$i'$) and ($g'$–$r'$) lying within ±0.15 and ±0.3 from the median values of $r'$–$i'$=0.47 and $g'$–$r'$=1.18 colours of the spectroscopically cluster members (see Fig. 6).

Figure 7 shows the contour map for the 579 likely cluster members having $r' \leq 21$. The 2D DEDICA procedure finds for significant peaks. In order of relative density, they are the E peak, the NW peak, the SW peak and the N peak. For each of these subclumps, Table 2 lists the number of assigned members, $N_S$; the peak position; the density (relative to the densest peak), $\rho_S$; the $\chi_S^2$ value. Ramella et al. (2007) tested the 2D DEDICA procedure with Monte–Carlo simulations reproducing galaxy clusters. They show that the physical significance associated to the subclusters is based on the statistical significance of the subcluster (recovered from the $\chi^2$ value) and the $r_{CS} = N_C/N_S$ parameter, where $N_S$ is the number of members of the substructure and $N_C + N_S$ is the total number of cluster members. The peaks listed in Table 2 have $r_{CS}$ in the range of $4 - 6$ and, as expected for real comparable substructures, $\chi^2 \gtrsim 25$ (see Fig. 2 of Ramella et al. 2007). We can estimate that these four subclumps have a small probability to be false detections (< 5%). Similar results are found for samples based on more conservative magnitude limits (e.g. $r' \leq 20.5$ and $r' \leq 20$).

To furtherly probe the robustness of these detections, we also apply the Voronoi Tessellation and Percolation (VTP) technique (e.g. Ramella et al. 2001; Barrena et al. 2005). This technique is non–parametric and does not smooth the data. As a consequence, it identifies galaxy structures irrespective of their shapes. For our purposes we run VTP on the same sample of 579 likely members as above. The result of the application of VTP is shown in Fig. 8. VTP is run three times adopting three detection thresholds: galaxies identified as belonging to structures at 95%, 98% and 99% c.ls. are shown as open squares, asterisks and solid circles respectively. VTP confirms the existence of the E, NW and SW clumps found with DEDICA, but doesn't find any significant structure in the position of the N clump. Since the existence of this clump is not confirmed we ignore it in the following. Instead, ~9′ south of the BCG, VTP detects a small galaxy peak, possibly a poor group infalling in the potential well of A2345.

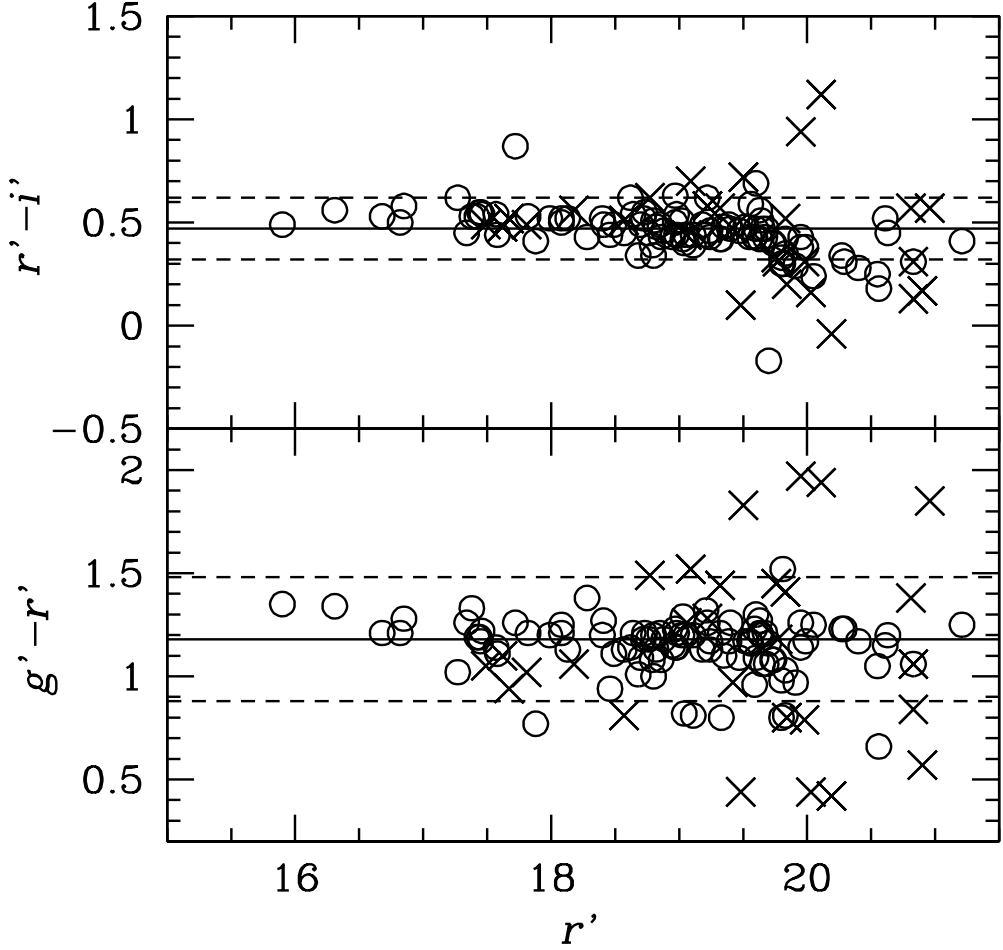

**Fig. 6.** *Upper panel: r′–i′* vs. *r′* diagram for galaxies with available spectroscopy. Circles and crosses indicate cluster members and non members. The solid line gives the median colour determined on member galaxies; the dashed lines are drawn at ±0.15 mag from this value. *Lower panel: g′–r′* vs. *r′* diagram for galaxies with available spectroscopy. The dashed lines are drawn at ±0.3 mag from the median value determined on member galaxies.

We also use a ROSAT/HRI archival image (pointing US800769H) to study the 2D structure of the cluster. The image has an exposure time of ~12.4 ks. As already noticed by D02, these data show a very large amount of substructure. X–ray contours in Fig. 1 show as the cluster structure is elongated in the E–W direction and the likely presence of three subclumps. The wavelet multiscale analysis performed on this image confirms the significance of the two most eastern subclumps, centered at R.A.=$21^h27^m13\overset{s}{.}99$, Dec.=$-12°09′53.3″$ (J2000.0; close to the BCG position) and R.A.=$21^h27^m05\overset{s}{.}98$, Dec.=$-12°10′44.3″$ (J2000.0), respectively. The centers of these two wavelet structures, as well as a rough center indicator for the most western X–ray peak, are compared to the position of galaxy subclumps in Fig. 7.

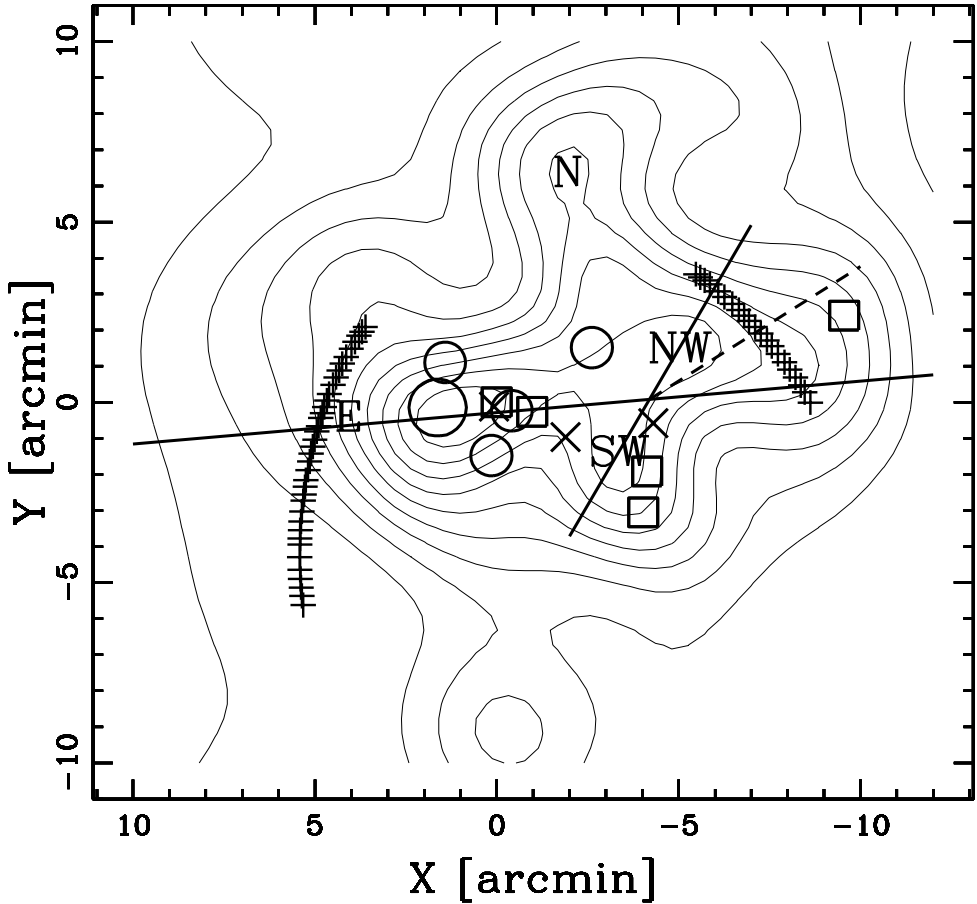

**Fig. 7.** Spatial distribution on the sky and relative isodensity contour map of likely cluster members with $r' \leq 21$, obtained with the 2D DEDICA method (external regions are not shown for the sake of clarity). The BCG is taken as the cluster center. The location of the five brightest galaxies are indicated by squares. Crosses indicate the peaks found in the ROSAT/HRI image. The circles indicate the position of the main (the largest circle) and the secondary peaks in the projected mass distribution as obtained from gravitational lensing analysis (see D02, notice that they analyzed only a small region around the BCG). The two relics are indicated in a schematic way by the "+" symbols. Solid lines highlight the directions of the two mergers. The dashed line is the bisecting of the merger directions (see text).

### 3.5. 3D–analysis

The existence of correlations between positions and velocities of cluster galaxies is a footprint of real substructures. Here we use several different approaches to analyze the structure of A2345 combining position and velocity information of the spectroscopic sample.

We analyze the presence of a velocity gradient performing a multiple linear regression fit to the observed velocities with respect to the galaxy positions in the plane of the sky and perform 1000 Monte Carlo simulations to assess the significance of this velocity gradient (e.g, Boschin et al. 2004 and refs. therein). We find no significant velocity gradient.

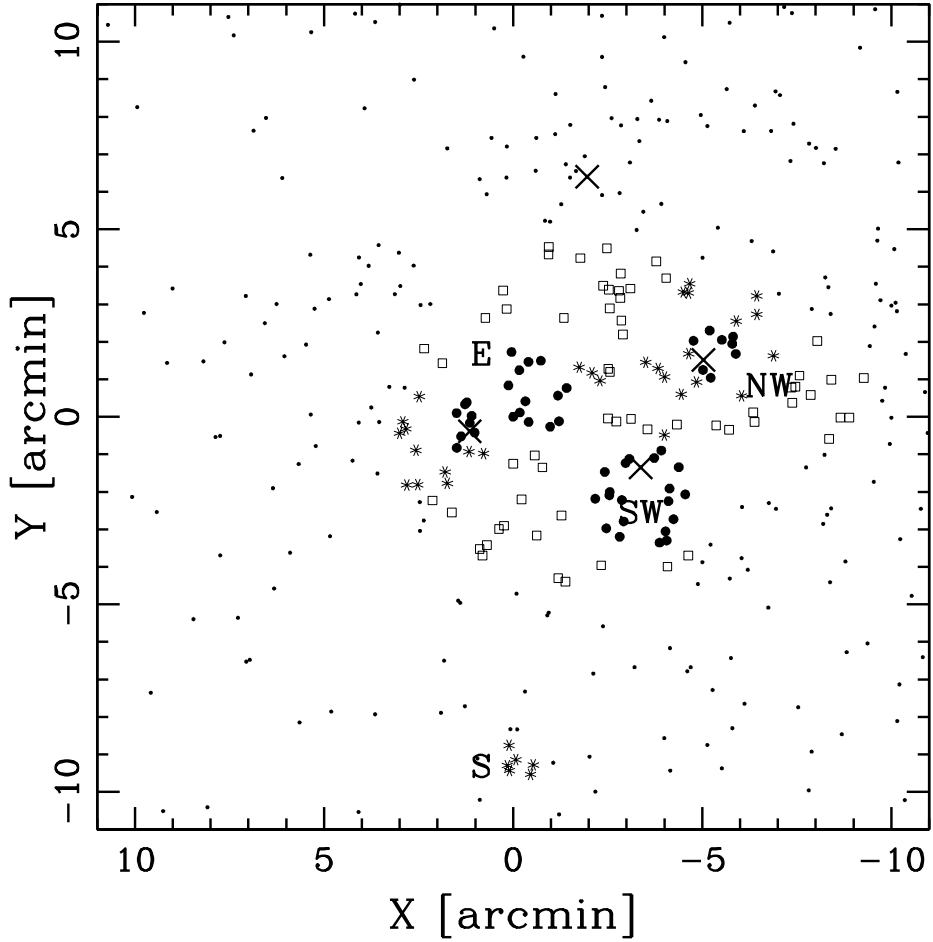

**Fig. 8.** Galaxies belonging to structures as detected by the Voronoi Tessellation and Percolation technique. The algorithm is run on the sample of likely members with $r' < 21$ (see text). Open squares, asterisks and solid circles indicate galaxies in structures at the 95%, 98% and 99% c. ls., respectively. The E, NW and SW subclumps and the S group are indicated. As a comparison, big crosses indicate the positions of the subclumps found with DEDICA.

We also combine galaxy velocity and position information to compute the $\Delta$–statistics devised by Dressler & Schectman (1988; see also Boschin et al. 2006 for a recent application) which is very useful to detect small, compact substructures. We find no significant indication of substructure.

The spectroscopic sample covers only the regions of the E, SW, and NW peaks detected in the galaxy density distribution (see the above section). To study these regions we select galaxies within 0.3 $h_{70}^{-1}$ Mpc from each peak, where the limiting radius is chosen to avoid any member superposition, and obtain three subclumps containing 18, 8 and 9 galaxies respectively. According to the standard means–test (e.g., Press et al. 1992) the E and NW clumps have different mean velocities at the 93% c.l., with the E clump having a larger velocity than the NW clump [$\langle v_E \rangle$ = (54573±232) km s$^{-1}$; $\langle v_{NW} \rangle$ = (53487±407) km s$^{-1}$]. We obtain a similar result comparing the E clump with the combined sample of SW+NW clumps [$\langle v_{NW+SW} \rangle$ = (53713±300) km s$^{-1}$].

The existence of a velocity difference between subclumps detected in two–dimensions prompted us to use the 3D KMM analysis to properly assign galaxies to the three subclumps, although the use of a Gaussian for the galaxy distribution is very approximate. We apply the 3D KMM algorithm both on the whole cluster sample and the subsample of 88 galaxies having $r' < 20$, i.e. with a relevant level of magnitude completeness. In both cases the algorithm fits a three–group partition, at the 99.4% c.l. (at the 98.4% c.l.) according to the likelihood ratio test, leading to three groups of 52, 17 and 29 galaxies (45, 13 and 30 galaxies), corresponding to the E, SW, and NW clumps. The results for the three KMM groups coming from both the analyses are shown in Table 3 (see also Figs. 5 and 9 for the $r' < 20$ sample).

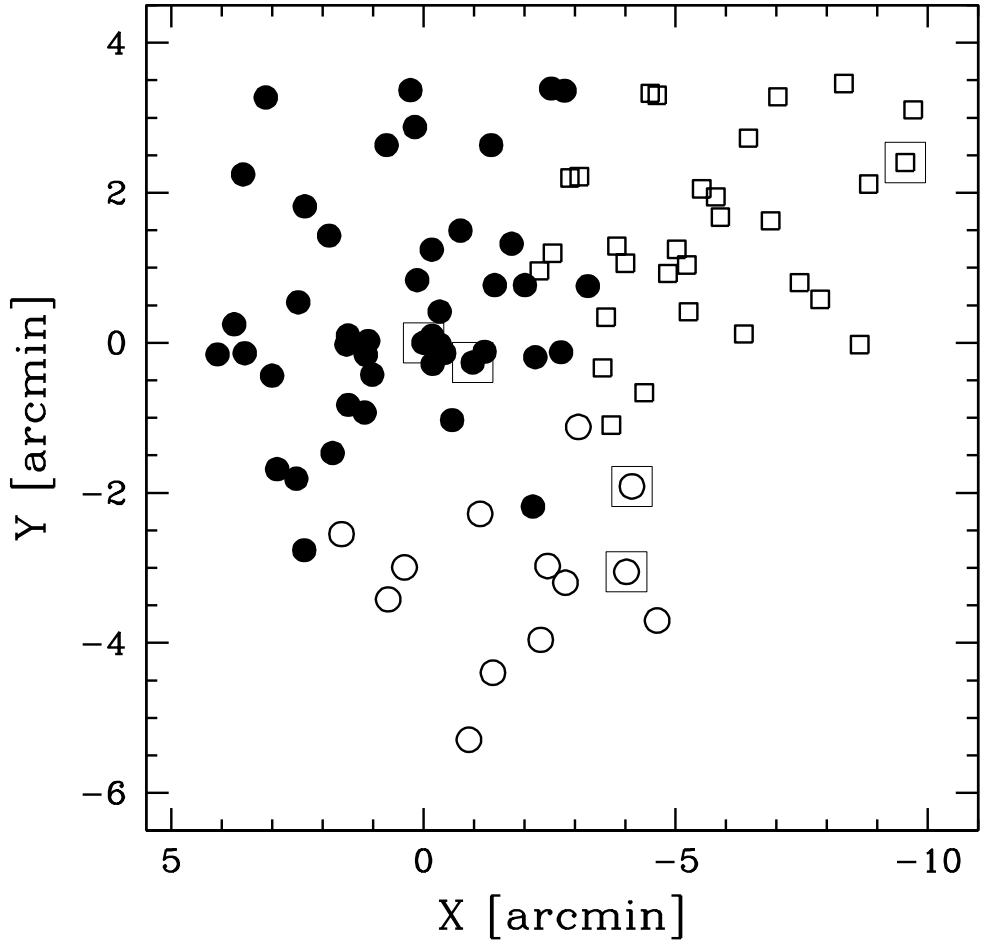

**Fig. 9.** Spatial distribution on the sky of the cluster galaxies (with $r' < 20$) showing the three groups recovered by the 3D KMM analysis. Solid circles, open circles and squares indicate the galaxies of E, SW and NW group, respectively. The BCG is taken as the cluster center. Large squares indicate the five brightest cluster members.

## 4. Discussion and conclusions

Our estimate of the cluster redshift is $\langle z \rangle = 0.1789 \pm 0.0004$ (cfr. with $z = 0.176$ reported in the literature as estimated by only one galaxy).

For the first time the internal dynamics of A2345 is analyzed on the basis of member galaxies.

The global value of the velocity dispersion $\sigma_V = 1069^{+75}_{-62}$ km s$^{-1}$ corresponds to that of a very rich cluster (e.g. the Coma cluster, Colless et al 1996; Girardi et al. 1998) although it is not particularly high within the DARC sample. The position of A2345 in the $L_{X,\text{bol}}$–$\sigma_V$ plane is well consistent with that of other clusters (see Fig. 5 of Ortiz–Gil et al. 2004, taking into account the different cosmologies).

### 4.1. Cluster structure

Our analysis of photometric data shows the presence of three significant peaks in the galaxy distribution, i.e. the E, SW, and NW peaks in order of importance. The E–peak is off–set to the east from the BCG by ∼ 1.3′ ($\gtrsim 0.2\,h_{70}^{-1}$ Mpc).

The presence of E, SW, and NW galaxy groups is also confirmed including our spectroscopic data in the analysis. We use different techniques to assign galaxies to the corresponding three subgroups. The corresponding kinematical properties are shown in Table 3. The E group is found to have the largest mean velocity $\langle v_E \rangle \gtrsim 54000$ km s$^{-1}$, while the SW and NW groups have similar mean velocities with a rest frame difference of $\Delta V_{\text{rf}} \lesssim 800$ km s$^{-1}$ with respect to the E group. The combination of these groups produces the double–peak structure found by our initial analysis of the spectroscopic sample (cfr. the three Gaussians with the position of the two 1D DEDICA peaks in Fig. 5).

The complexity of A2345 structure and the small velocity differences among subgroups are the likely cause of the failure of global, standard 3D techniques when searching for substructures (e.g. the velocity gradient and the Dressler–Schectman statistics). Our results suggest that the SW and NW clumps define a direction parallel to the plane of sky, while the relative direction between the E clump and the SW+NW complex have a component along the LOS.

### 4.2. Cluster mass

The complicated structure of A2345 and the small velocity differences among subgroups also make us difficult to estimate the velocity dispersions for the individual subclumps. Table 3 lists what we nominally obtain in our analysis but, while the values of $\sigma_V$ computed in the 1D DEDICA analysis are obvious underestimates of the true values due to the abrupt division of the velocity distribution, the values computed using other techniques are likely overestimates due to the possible contamination from the other, close groups. The 3D KMM method suggests that the group with the largest $\sigma_V$, i.e. the most massive one, is the E group. This also agrees with the fact that the E group coincides with a strong mass peak (see D02) and hosts the BCG and the main X–ray peak. However, the precise value of $\sigma_{V,E}$ is not obvious: Fig. 4 shows as the galaxy group around the BCG has a $\sigma_V \sim 800$ km s$^{-1}$ increasing out to $\sigma_V \gtrsim 1100$ km s$^{-1}$ when considering more and more distant (and so likely more contaminated) regions. This range of $\sigma_{V,E}$ values overlaps with that of the $\sigma_{WL}$ values obtained by the weak gravitational lensing analysis for a singular isothermal sphere model or other approaches (870 − 965 km s$^{-1}$, D02; Cypriano et al. 2004). Hereafter, we will assume that $\sigma_{V,E} \sim 900$ km s$^{-1}$ for the E group.

The uncertainties are even larger for the other two groups. The relative density in the 2D galaxy distribution, the amount of the population obtained through the 3D DEDICA method and the mass map (from the gravitational lensing, see the section below) suggest that the NW group is more massive than the SW group, but the values of $\sigma_V$ are not conclusive. We decide to assume for both the groups $\sigma_{V,SW} = \sigma_{V,NW} \sim 700$ km s$^{-1}$, a value somewhat smaller than that for the E group.

Finally, we attempt to estimate the mass of the whole A2345 complex. Following the methodology for the virial mass estimate already outlined in previous papers for other clusters of DARC sample (e.g., Girardi et al. 1998; Girardi & Mezzetti 2001; Boschin et al. 2009), the assumption of the dynamical equilibrium for a cluster having $\sigma_V = 1069$ km s$^{-1}$ would lead to a radius of the quasi–virialized region $R_{vir} = 0.17 \times \sigma_V/H(z) = 2.4\ h_{70}^{-1}$ Mpc and a mass of $M(< R \sim 2.4\ h_{70}^{-1} \text{Mpc}) \sim 1.8 \times 10^{15}\ h_{70}^{-1}\ M_\odot$. Considering the combination of the three above subclumps, each assumed to be in equilibrium, we instead obtain $M \sim 1.1 + 0.5 + 0.5 \sim 2.1 \times 10^{15}\ h_{70}^{-1}\ M_\odot$.

### 4.3. Merging phase

To investigate the phase of the cluster merger among the detected galaxy clumps we must resort to the comparison with results from other wavelengths. The complex cluster structure of A2345 is also revealed using X–ray and gravitational lensing analysis. The analysis of ROSAT/HRI data shows the presence of two, likely three, peaks in the X–ray image, with two X–ray peaks lying between E and SW peaks (of which one coincident with the BCG) and one X–ray peak lying between SW and NW peaks. No peak is shown by the HRI image in the northern region reinforcing the global E–W (or better ENE–WSW) elongation of the X–ray surface brightness. Two more peaks are shown by ROSAT/PSPC archival data in the most external NW region, in particular the peak "X1" (see B09) is close to the position of the second brightest galaxy in our spectroscopic catalog (ID. 2).

As for the weak gravitational lensing approach, D02 analyze a small cluster region of 6′×6′, i.e. in practice the region around our E group. The mass map shows a main peak off–set to the east from the BCG by $\sim 1.5′$, but well coincident with the position of the E peak as found in the 2D galaxy distribution. Unfortunately, both SW and NW peaks lie outside the region analyzed by D02. However, very interestingly, the mass map shows a strong elongation toward the NW region in the direction of the NW peak.

The displacement between X–ray and optical (or mass) peaks suggests that A2345 is in an "after cross core" phase of merger. The observational phenomenology indicate two merging directions. The X–ray brightness distribution is mainly elongated towards the E–W (slightly ENE–WSW) direction, the same direction defined by the E peak and the other two galaxy peaks (SW+NW peaks). This direction is roughly perpendicular to the elongation of the (eastern) A2345–2 radio relic. The second direction is a S–N (slightly SSE–NNW) direction as indicated by the SW and NW peaks, and the intermediate X–ray peak. The bisecting line of this and the above merging direction is roughly perpendicular to the direction of the elongation of the (western) A2345–1 radio relic (see Fig. 7). The same bisecting direction is also roughly indicated by the tail of the radio galaxy ID. 35 (Owen & Ledlow 1997, B09).

In the framework of the "outgoing merger shocks" model, the E–W merger is the natural origin of the formation of the eastern radio relic (A2345–2), while the formation of the expected western radio relic was disturbed by the presence of the additional S–N merger in the western cluster regions resulting in the formation of the north–western A2345–1 radio relic. This scenario would explain both the asymmetry of the two relics and, qualitatively, the peculiarity of A2345–1. In fact, the morphology of A2345–1 is different from the usual arc–like morphology with the presence of a kind of tail departing from the South of the relic and bended towards the external regions (see left–upper panel of Fig. 2 of B09). Moreover, the southern and external regions of A2345–1 show a higher spectral index than the northern region suggesting that the northern region has received an additional input of energy.

A difficulty in our above scenario is the absence of a southern relic, which we would expect due to the merger of the SW group with the NW group. However, this could be connected with the fact that the southern group is likely the less important galaxy clump in A2345.

### 4.4. Toy model for the complex merger

In the framework of the observational picture and the above scenario we attempt to quantify the internal cluster dynamics of A2345 through a simple analytical approach.

In the case of a two–groups merger, the "outgoing merger shocks" model is successfully supported by the kinematical data on subclumps through the use of the simple analytical bimodal model (e.g. Barrena et al. 2009 for Abell 1240).

As for A2345, we assume a toy model, obviously very approximate, where the cluster is first interested by the merger between the E group and a W group (formed by the SW+NW groups). Then, the western part of the cluster is interested by a second merger between the SW group and the NW group. In practice, according to this toy model, the internal dynamics of A2345 is analyzed by the combination of two successive bimodal models. For both these mergers we apply the two–body model (Beers et al. 1992; Thompson 1982) following the methodology outlined for A1240 (Barrena et a. 2009).

The two–body model assumes radial orbits for the clumps with no shear or net rotation of the system. According to the boundary conditions usually considered, the clumps are assumed to begin their evolution at time $t_0 = 0$ with a separation $d_0 = 0$, and are now moving apart or coming together for the first time in their history. In the case of a collision, we assume that the time $t_0 = 0$ with separation $d_0 = 0$ is the time of their core crossing and that we are looking at the system a time $t$ after.

As for the first collision, the values of relevant parameters for the two–clump system (the E group and the SW+NW complex) are the relative LOS velocity in the rest–frame, $V_{\rm rf} \sim 800$ km s$^{-1}$, and the projected linear distance between the two clumps, $D \sim 1\ h_{70}^{-1}$ Mpc. To obtain an estimate of $t$, we use the Mach number of the shock $\mathcal{M} \sim 2.8$ as inferred by B09 from the radio spectral index of A2345–2. The Mach number is defined to be $\mathcal{M} = v_{\rm s}/c_{\rm s}$, where $v_{\rm s}$ is the velocity of the shock and $c_{\rm s}$ is the sound speed in the pre–shock gas (see e.g., Sarazin 2002 for a review). The value of $c_{\rm s}$, obtained from our estimate of $\sigma_{\rm V,E} \sim 900$ km s$^{-1}$, leads to a value of $v_{\rm s} \sim 2.5 \times 10^3$ km s$^{-1}$. Assuming the shock velocity to be a constant, the shock covered a $\sim 0.9\ h_{70}^{-1}$ Mpc scale (i.e., the distance of the relic from the cluster center) in a time of $\sim 0.35$ Gyrs. We assume this time as our estimate of $t$. Although the velocity of the shock is not constant, studies based on numerical simulations show how the variation in $v_{\rm s}$ is much smaller than the variation in the relative velocity of the subclumps identified with their dark matter components (see Fig. 4 of Springel & Farrar 2007 and Fig. 14 of Mastropietro & Burkert 2008), thus our rough estimate of $t$ is acceptable as a first order approximation.

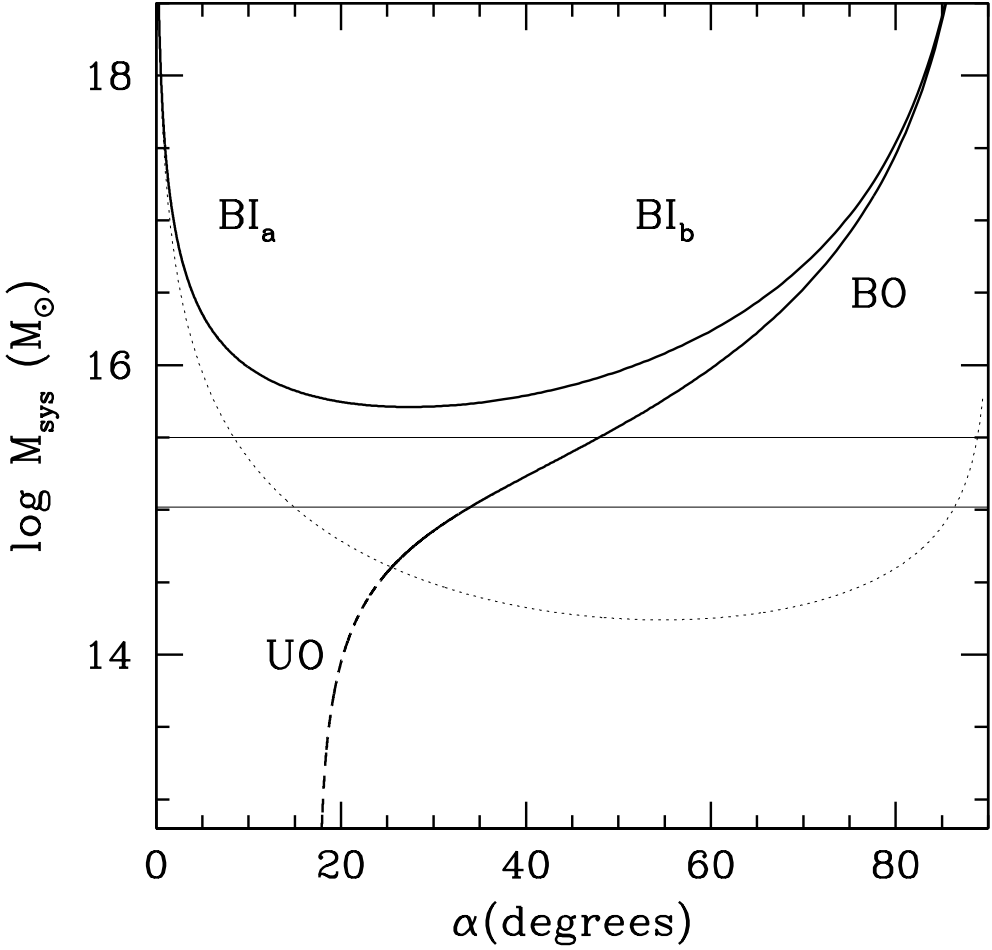

**Fig. 10.** System mass vs. projection angle for bound and unbound solutions (thick solid and thick dashed curves, respectively) of the two–body model applied to the E and SN+SW subsystems. Labels BI$_a$ and BI$_b$ indicate the bound and incoming, i.e., collapsing solutions (thick solid curve). Label BO indicates the bound outgoing, i.e., expanding solutions (thick solid curve). Label UO indicates the unbound outgoing solutions (thick dashed curve). The horizontal lines give the range of observational values of the mass system with a 50% error. The thin dashed curve separates bound and unbound regions according to the Newtonian criterion (above and below the thin dashed curve, respectively).

The bimodal model solution gives the total system mass $M_{sys}$, i.e. the sum of the masses of the E+SW+NW groups, as a function of $\alpha$, where $\alpha$ is the projection angle between the plane of the sky and the line connecting the centers of the two clumps (e.g., Gregory & Thompson 1984). Figure 10 compares the bimodal–model solutions with the observed mass of the system considering a 50% uncertainty band. The present solutions span the bound outgoing solutions (i.e., expanding), BO; the bound incoming solutions (i.e., collapsing), BI$_a$ and BI$_b$; and the unbound outgoing solutions, UO. For the incoming case, there are two solutions because of the ambiguity in the projection angle $\alpha$. The BO solution is the only one to be consistent with the observed mass range leading to a $\alpha$ ~40–50 degrees. The BO solution means that the E group is moving towards East going in the opposite direction with respect to the observer, while the SW+NW complex is moving toward West toward the observer. The angle estimate means that the true spatial distance between the two subclumps is $\sim 1.3$–$1.5$ $h_{70}^{-1}$ Mpc and that the real, i.e. deprojected, velocity difference is $V_{\rm rf} \sim 1000 - 1250$ km s$^{-1}$. We note that the present relative velocity between galaxy clumps is smaller than the shock velocity, i.e., the regime is not stationary, but this is expected when comparing shock and collisionless components in numerical simulations (Springel & Farrar 2007; Mastropietro & Burkert 2008). The deprojected velocity difference of the two clumps at the cross core time is $V_{\rm rf} \sim 2000 - 2800$ km s$^{-1}$, i.e. comparable to that of the shock as expected.

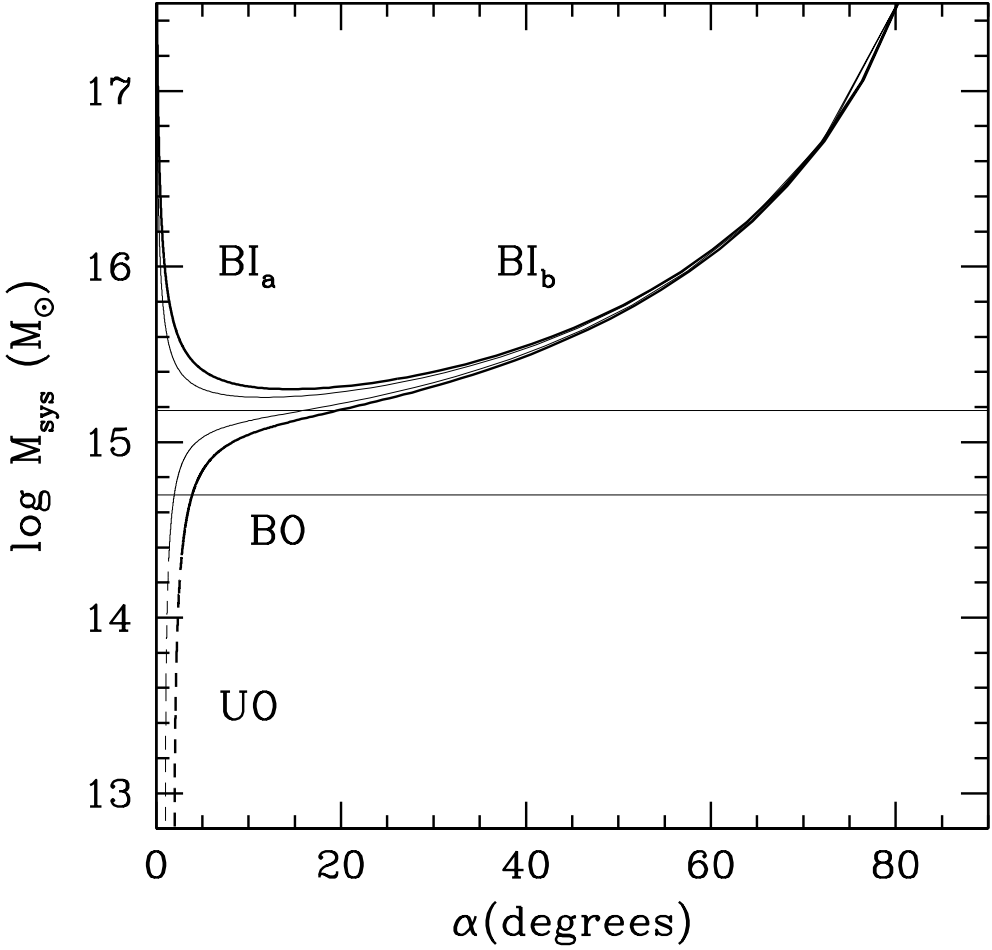

**Fig. 11.** The same that in Fig. 10 but for the NW and SW subsystems. Thick and thin lines give the results for $V_{\rm rf,LOS} \sim 100$ and 50 km s$^{-1}$, respectively.

As for the second collision, the values of relevant parameters for the two–clump system (the SW and the NW groups) are $V_{\rm rf} \sim 0$ km s$^{-1}$ (since we do not see any significant velocity difference), $D \sim 0.3$ $h_{70}^{-1}$ Mpc and $M_{\rm sys} \sim 1.1 + 0.5 \sim 1.6 \times 10^{15}$ $h_{70}^{-1}$ $M_\odot$. In our scenario this merger is more recent than the above one and thus we assume a time $t \sim 0.2$ Gyrs.

Figure 11 shows the results for $V_{\rm rf} = 50$ km s$^{-1}$ and $V_{\rm rf} = 100$ km s$^{-1}$. The BO solution is well acceptable with $\alpha \sim 2 - 20°$. This means that the SW group is moving toward South and the NW group towards North and, as someway expected from the observations, the NW and SW groups define a direction almost parallel to the plane of the sky. The true spatial distance is similar to the projected one. The real velocity difference has a huge uncertainty due to its dependence from the very uncertain $\alpha$ value. Notice that this merger should be considered of minor importance with respect to the main one described above, thus here the bimodal model is likely a worse representation of the reality than the above case.

In conclusion, we recover from our toy model quantitative results consistent with the global scenario. Thus, our analysis of the internal cluster dynamics well supports the "outgoing merger shocks" model proposed by B09. We also present an explanation for the relics asymmetry and the peculiarity of A2345–1. Our explanation is based on the internal structure of A2345, rather than on a possible merger with an external, close group (e.g. the group "X1" detected by B09). Before applying more reliable approaches to the study of A2345 internal dynamics, an improvement of the present observational picture would be very useful, e.g. by obtaining redshift information for more galaxies and deeper X–ray data (e.g. with Chandra and/or XMM telescopes) to compute gas temperature maps.

*Acknowledgements.* We are in debt with Annalisa Bonafede for the VLA radio image she kindly provided us. We thank the anonymous referee for his/her useful comments and suggestions. This publication is based on observations made on the island of La Palma with the Italian Telescopio Nazionale Galileo (TNG) and the Isaac Newton Telescope (INT). The TNG is operated by the Fundación Galileo Galilei – INAF (Istituto Nazionale di Astrofisica). The INT is operated by the Isaac Newton Group. Both telescopes are located in the Spanish Observatorio of the Roque de Los Muchachos of the Instituto de Astrofísica de Canarias.

This research has made use of the NASA/IPAC Extragalactic Database (NED), which is operated by the Jet Propulsion Laboratory, California Institute of Technology, under contract with the National Aeronautics and Space Administration.

## List of Objects

**Table 1.** Velocity catalog of 125 spectroscopically measured galaxies in the field of the cluster A2345. ID. 82, in boldface, highlights the BCG.

| ID | $\alpha, \delta$ (J2000) | $g'$ | $r'$ | $i'$ | v (km s$^{-1}$) | $\Delta$v | Cl. |
|---|---|---|---|---|---|---|---|
| 1 | 21 26 33.96, −12 06 39.6 | 20.90 | 19.63 | 19.07 | 53195 | 50 | 1 |
| 2 | 21 26 34.61, −12 07 21.7 | 17.65 | 16.31 | 15.75 | 53434 | 42 | 1 |
| 3 | 21 26 35.76, −12 08 43.8 | 18.72 | 17.62 | 17.14 | 49692 | 29 | 2 |
| 4 | 21 26 36.84, −12 09 13.7 | 21.57 | 20.40 | 20.12 | 54606 | 78 | 1 |
| 5 | 21 26 37.58, −12 07 39.0 | 20.86 | 19.70 | 19.87 | 54760 | 77 | 1 |
| 6 | 21 26 38.30, −12 09 47.5 | 20.74 | 19.68 | 19.21 | 54131 | 52 | 1 |
| 7 | 21 26 39.60, −12 06 18.7 | 18.60 | 17.34 | 16.89 | 52880 | 38 | 1 |
| 8 | 21 26 40.80, −12 07 45.1 | 21.76 | 20.61 | 20.09 | 52719 | 81 | 1 |
| 9 | 21 26 41.50, −12 09 11.2 | 18.71 | 17.57 | 17.03 | 52238 | 41 | 1 |
| 10 | 21 26 42.58, −12 10 04.4 | 22.81 | 20.96 | 20.39 | 109606 | 63 | 2 |
| 11 | 21 26 43.18, −12 08 57.8 | 18.62 | 17.45 | 16.90 | 54077 | 48 | 1 |
| 12 | 21 26 44.06, −12 10 05.2 | 21.33 | 19.50 | 18.78 | 109806 | 42 | 2 |
| 13 | 21 26 44.95, −12 06 29.2 | 20.69 | 19.52 | 19.04 | 52462 | 48 | 1 |
| 14 | 21 26 45.53, −12 08 08.5 | 19.95 | 18.82 | 18.34 | 54222 | 39 | 1 |
| 15 | 21 26 47.33, −12 07 02.3 | 18.68 | 17.46 | 16.92 | 54033 | 45 | 1 |
| 16 | 21 26 47.71, −12 09 38.9 | 21.16 | 19.99 | 19.61 | 53128 | 48 | 1 |
| 17 | 21 26 48.50, −12 05 48.0 | 20.32 | 19.06 | 18.62 | 50013 | 56 | 2 |
| 18 | 21 26 48.94, −12 12 10.4 | 20.50 | 19.22 | 18.64 | 78160 | 41 | 2 |
| 19 | 21 26 49.61, −12 08 05.6 | 20.22 | 19.02 | 18.51 | 55321 | 56 | 1 |
| 20 | 21 26 49.82, −12 11 38.4 | 19.92 | 19.48 | 19.38 | 18109 | 67 | 2 |
| 21 | 21 26 49.99, −12 07 49.4 | 19.60 | 18.49 | 18.00 | 53224 | 35 | 1 |
| 22 | 21 26 51.12, −12 07 43.0 | 20.86 | 19.83 | 19.41 | 51907 | 73 | 1 |
| 23 | 21 26 51.46, −12 12 59.0 | 21.60 | 20.55 | 20.30 | 54728 | 84 | 1 |
| 24 | 21 26 52.18, −12 09 21.2 | 19.86 | 19.04 | 18.64 | 53423 | 83 | 1 |
| 25 | 21 26 52.32, −12 08 43.8 | 19.87 | 18.79 | 18.40 | 54883 | 31 | 1 |
| 26 | 21 26 53.16, −12 08 31.2 | 19.40 | 18.46 | 18.02 | 53500 | 53 | 1 |
| 27 | 21 26 53.30, −12 10 27.5 | 22.05 | 20.11 | 18.99 | 135483 | 64 | 2 |
| 28 | 21 26 53.86, −12 08 50.6 | 20.71 | 19.65 | 19.18 | 52016 | 48 | 1 |
| 29 | 21 26 54.74, −12 06 28.1 | 20.10 | 18.95 | 18.52 | 53441 | 48 | 1 |
| 30 | 21 26 54.74, −12 13 28.2 | 19.26 | 18.13 | 17.61 | 53156 | 46 | 1 |
| 31 | 21 26 55.30, −12 06 26.6 | 19.90 | 18.72 | 18.20 | 53109 | 39 | 1 |
| 32 | 21 26 55.78, −12 10 26.0 | 20.64 | 19.83 | 19.48 | 52690 | 71 | 1 |
| 33 | 21 26 56.78, −12 11 40.9 | 18.03 | 16.82 | 16.32 | 52257 | 36 | 1 |
| 34 | 21 26 56.86, −12 07 23.9 | 21.47 | 20.90 | 20.73 | 69905 | 122 | 2 |
| 35 | 21 26 57.22, −12 12 49.3 | 17.89 | 16.68 | 16.15 | 52962 | 35 | 1 |
| 36 | 21 26 57.31, −12 08 42.4 | 19.19 | 17.99 | 17.47 | 54590 | 38 | 1 |
| 37 | 21 26 58.01, −12 08 28.7 | 20.78 | 19.80 | 19.47 | 52616 | 116 | 1 |
| 38 | 21 26 58.44, −12 10 52.0 | 19.60 | 18.40 | 17.88 | 54442 | 38 | 1 |
| 39 | 21 26 58.87, −12 09 25.6 | 20.88 | 19.91 | 19.62 | 53371 | 57 | 1 |
| 40 | 21 26 59.16, −12 10 06.2 | 20.06 | 18.85 | 18.36 | 52785 | 60 | 1 |
| 41 | 21 27 00.36, −12 09 00.7 | 20.20 | 18.97 | 18.34 | 54778 | 49 | 1 |
| 42 | 21 27 01.06, −12 07 33.2 | 19.92 | 19.11 | 18.72 | 50847 | 38 | 1 |

**Table 1.** Continued.

| ID | $\alpha, \delta$ (J2000) | $g'$ | $r'$ | $i'$ | v | $\Delta$v | Cl. |
|---|---|---|---|---|---|---|---|
| | | | | | | ( km s$^{-1}$ ) | |
| 43 | 21 27 01.13, −12 10 53.4 | 18.61 | 17.42 | 16.89 | 53044 | 36 | 1 |
| 44 | 21 27 01.75, −12 12 33.5 | 18.61 | 17.67 | 17.15 | 37854 | 56 | 2 |
| 45 | 21 27 01.82, −12 07 34.3 | 19.94 | 18.73 | 18.18 | 52305 | 45 | 1 |
| 46 | 21 27 02.18, −12 12 58.0 | 19.29 | 18.08 | 17.58 | 54730 | 28 | 1 |
| 47 | 21 27 02.26, −12 06 24.5 | 20.19 | 18.98 | 18.54 | 54621 | 42 | 1 |
| 48 | 21 27 02.54, −12 09 53.6 | 19.68 | 18.41 | 17.92 | 56034 | 50 | 1 |
| 49 | 21 27 03.22, −12 08 34.4 | 20.12 | 18.98 | 18.44 | 53128 | 49 | 1 |
| 50 | 21 27 03.34, −12 06 22.7 | 20.82 | 19.59 | 19.16 | 54660 | 41 | 1 |
| 51 | 21 27 03.62, −12 12 44.6 | 19.85 | 18.64 | 18.10 | 53909 | 45 | 1 |
| 52 | 21 27 04.18, −12 13 43.7 | 19.94 | 18.76 | 18.31 | 54267 | 28 | 1 |
| 53 | 21 27 04.30, −12 08 48.5 | 20.82 | 19.74 | 19.34 | 52661 | 43 | 1 |
| 54 | 21 27 04.63, −12 09 57.6 | 18.98 | 17.72 | 16.85 | 55154 | 50 | 1 |
| 55 | 21 27 04.82, −12 11 57.1 | 20.53 | 19.21 | 18.77 | 55742 | 50 | 1 |
| 56 | 21 27 05.47, −12 09 00.0 | 20.69 | 19.60 | 18.91 | 52837 | 55 | 1 |
| 57 | 21 27 06.24, −12 11 48.1 | 21.67 | 20.83 | 20.70 | 37665 | 92 | 2 |
| 58 | 21 27 06.55, −12 08 26.9 | 19.94 | 18.86 | 18.43 | 55026 | 37 | 1 |
| 59 | 21 27 07.10, −12 08 25.8 | 20.77 | 19.98 | 19.67 | 57591 | 84 | 2 |
| 60 | 21 27 07.20, −12 14 14.3 | 20.95 | 19.77 | 19.46 | 38060 | 67 | 2 |
| 61 | 21 27 07.92, −12 09 00.0 | 20.84 | 19.64 | 19.13 | 53009 | 63 | 1 |
| 62 | 21 27 08.06, −12 14 10.0 | 18.29 | 17.27 | 16.65 | 53740 | 38 | 1 |
| 63 | 21 27 08.21, −12 07 08.0 | 20.84 | 19.63 | 19.21 | 55740 | 50 | 1 |
| 64 | 21 27 08.26, −12 10 27.1 | 21.92 | 19.95 | 19.01 | 148206 | 56 | 2 |
| 65 | 21 27 08.47, −12 12 24.1 | 19.24 | 18.18 | 17.62 | 38450 | 60 | 2 |
| 66 | 21 27 08.74, −12 09 53.3 | 20.90 | 19.60 | 19.13 | 52382 | 56 | 1 |
| 67 | 21 27 08.86, −12 09 12.2 | 21.50 | 20.27 | 19.93 | 52861 | 60 | 1 |
| 68 | 21 27 08.98, −12 08 43.8 | 21.22 | 20.56 | 20.38 | 51038 | 101 | 1 |
| 69 | 21 27 09.10, −12 12 02.9 | 18.65 | 17.88 | 17.47 | 52080 | 36 | 1 |
| 70 | 21 27 09.70, −12 10 01.9 | 18.13 | 16.85 | 16.27 | 53716 | 45 | 1 |
| 71 | 21 27 10.01, −12 15 03.6 | 20.48 | 19.22 | 18.60 | 52337 | 39 | 1 |
| 72 | 21 27 10.39, −12 11 23.3 | 19.38 | 18.57 | 18.06 | 38093 | 49 | 2 |
| 73 | 21 27 10.54, −12 11 07.1 | 20.39 | 19.42 | 18.95 | 38165 | 42 | 2 |
| 74 | 21 27 10.70, −12 08 16.4 | 19.69 | 18.68 | 18.34 | 55500 | 42 | 1 |
| 75 | 21 27 11.38, −12 10 48.0 | 20.41 | 19.24 | 18.81 | 53708 | 57 | 1 |
| 76 | 21 27 12.02, −12 09 54.4 | 18.71 | 17.38 | 16.85 | 53996 | 41 | 1 |
| 77 | 21 27 12.38, −12 09 21.2 | 19.66 | 18.28 | 17.85 | 55539 | 50 | 1 |
| 78 | 21 27 12.38, −12 09 47.5 | 19.76 | 18.62 | 18.00 | 54778 | 52 | 1 |
| 79 | 21 27 12.96, −12 10 03.0 | 21.33 | 19.81 | 19.51 | 53979 | 77 | 1 |
| 80 | 21 27 12.99, −12 09 40.4 | 19.03 | 17.82 | 17.29 | 52311 | 36 | 1 |
| 81 | 21 27 13.03, −12 08 31.6 | 20.81 | 19.73 | 19.33 | 53145 | 99 | 1 |
| **82** | 21 27 13.70, −12 09 46.1 | 17.25 | 15.90 | 15.41 | 54140 | 33 | 1 |
| 83 | 21 27 14.21, −12 08 56.0 | 19.33 | 18.08 | 17.56 | 55766 | 55 | 1 |
| 84 | 21 27 14.40, −12 06 53.6 | 21.09 | 19.95 | 19.57 | 54017 | 48 | 1 |

**Table 1.** Continued.

| ID | $\alpha, \delta$ (J2000) | $g'$ | $r'$ | $i'$ | v | $\Delta$v ( km s$^{-1}$ ) | Cl. |
|---|---|---|---|---|---|---|---|
| 85 | 21 27 14.59, −12 10 25.0 | 21.30 | 20.05 | 19.81 | 54554 | 81 | 1 |
| 86 | 21 27 14.76, −12 06 24.1 | 20.37 | 19.24 | 18.78 | 51578 | 39 | 1 |
| 87 | 21 27 15.12, −12 08 54.6 | 20.26 | 18.77 | 18.15 | 78821 | 73 | 2 |
| 88 | 21 27 15.24, −12 12 45.7 | 20.11 | 18.97 | 18.47 | 53272 | 36 | 1 |
| 89 | 21 27 16.03, −12 13 05.2 | 22.46 | 21.21 | 20.80 | 54931 | 84 | 1 |
| 90 | 21 27 16.56, −12 13 11.3 | 20.45 | 19.35 | 18.87 | 51322 | 43 | 1 |
| 91 | 21 27 16.70, −12 07 08.0 | 18.69 | 17.58 | 17.14 | 52194 | 38 | 1 |
| 92 | 21 27 17.11, −12 09 38.9 | 21.52 | 20.29 | 19.98 | 52250 | 199 | 1 |
| 93 | 21 27 17.33, −12 13 17.4 | 18.83 | 17.81 | 17.32 | 38647 | 29 | 2 |
| 94 | 21 27 17.86, −12 10 11.6 | 20.56 | 19.47 | 19.01 | 52994 | 70 | 1 |
| 95 | 21 27 18.19, −12 09 44.6 | 20.32 | 19.03 | 18.61 | 54748 | 42 | 1 |
| 96 | 21 27 18.38, −12 09 55.8 | 19.70 | 18.57 | 18.12 | 55774 | 43 | 1 |
| 97 | 21 27 18.48, −12 10 41.9 | 20.31 | 19.11 | 18.67 | 54179 | 39 | 1 |
| 98 | 21 27 18.91, −12 09 26.3 | 21.83 | 20.63 | 20.18 | 55710 | 154 | 1 |
| 99 | 21 27 19.80, −12 10 35.8 | 20.72 | 19.56 | 18.97 | 54444 | 63 | 1 |
| 100 | 21 27 19.82, −12 09 40.3 | 21.21 | 19.95 | 19.52 | 54651 | 64 | 1 |
| 101 | 21 27 19.94, −12 09 47.5 | 20.60 | 19.80 | 19.50 | 51373 | 73 | 1 |
| 102 | 21 27 20.35, −12 12 19.1 | 19.79 | 18.70 | 18.21 | 52923 | 29 | 1 |
| 103 | 21 27 21.07, −12 11 14.3 | 20.66 | 19.40 | 18.93 | 54413 | 62 | 1 |
| 104 | 21 27 21.36, −12 08 20.4 | 20.06 | 18.88 | 18.42 | 54312 | 60 | 1 |
| 105 | 21 27 22.44, −12 12 00.0 | 22.19 | 20.81 | 20.24 | 78472 | 91 | 2 |
| 106 | 21 27 23.33, −12 07 57.0 | 20.53 | 19.32 | 18.90 | 53097 | 71 | 1 |
| 107 | 21 27 23.38, −12 12 32.0 | 20.31 | 19.18 | 18.69 | 54017 | 48 | 1 |
| 108 | 21 27 23.54, −12 07 46.9 | 20.47 | 20.03 | 19.87 | 33646 | 55 | 2 |
| 109 | 21 27 23.78, −12 12 48.6 | 18.54 | 17.49 | 17.00 | 38564 | 49 | 2 |
| 110 | 21 27 23.86, −12 09 13.7 | 20.72 | 19.55 | 19.12 | 54842 | 67 | 1 |
| 111 | 21 27 24.02, −12 11 34.8 | 20.00 | 18.80 | 18.29 | 54486 | 45 | 1 |
| 112 | 21 27 25.30, −12 10 05.5 | 21.24 | 19.83 | 19.31 | 78709 | 55 | 2 |
| 113 | 21 27 25.58, −12 11 27.2 | 20.13 | 19.33 | 18.91 | 52024 | 56 | 1 |
| 114 | 21 27 25.99, −12 10 12.4 | 20.56 | 19.39 | 18.90 | 53719 | 50 | 1 |
| 115 | 21 27 26.50, −12 06 29.9 | 20.55 | 19.59 | 19.16 | 52048 | 71 | 1 |
| 116 | 21 27 27.10, −12 08 58.2 | 21.21 | 19.76 | 19.43 | 74089 | 55 | 2 |
| 117 | 21 27 28.20, −12 09 54.4 | 18.63 | 17.44 | 16.89 | 51945 | 41 | 1 |
| 118 | 21 27 28.34, −12 07 31.4 | 19.80 | 18.80 | 18.46 | 53266 | 43 | 1 |
| 119 | 21 27 29.06, −12 09 31.3 | 20.88 | 19.67 | 19.24 | 54126 | 49 | 1 |
| 120 | 21 27 30.41, −12 09 55.4 | 20.25 | 19.05 | 18.61 | 56240 | 56 | 1 |
| 121 | 21 27 31.94, −12 10 43.3 | 21.89 | 20.83 | 20.52 | 53681 | 101 | 1 |
| 122 | 21 27 32.95, −12 08 03.1 | 20.61 | 20.19 | 20.23 | 29128 | 116 | 2 |
| 123 | 21 27 34.30, −12 09 32.8 | 20.64 | 19.84 | 19.64 | 16834 | 125 | 2 |
| 124 | 21 27 35.40, −12 10 14.5 | 20.61 | 19.09 | 18.39 | 99120 | 57 | 2 |
| 125 | 21 27 36.10, −12 07 50.5 | 20.76 | 19.32 | 18.75 | 78075 | 77 | 2 |

**Table 2.** Substructure from the photometric sample detected with DEDICA.

| Subclump | $N_S$ | $\alpha, \delta$ (J2000) | $\rho_S$ | $\chi^2_S$ |
|---|---|---|---|---|
| 2D – E | 106 | 212718.4 – 121009 | 1.000 | 35.0 |
| 2D – NW | 99 | 212653.1 – 120815 | 0.887 | 30.0 |
| 2D – SW | 86 | 212659.9 – 121107 | 0.876 | 24.4 |
| 2D – N[a] | 86 | 212705.7 – 120322 | 0.655 | 25.9 |

[a] This subclump is NOT confirmed by the VTP analysis.

**Table 3.** Kinematical properties of cluster and subclumps.

| Analysis | System | $N_g$ | $<v>$ km s$^{-1}$ | $\sigma_V$ km s$^{-1}$ |
|---|---|---|---|---|
|  | A2345 | 98 | $53645 \pm 108$ | $1069^{+75}_{-62}$ |
| 1D DEDICA | low vel. peak[a] | 53 | 53202 | > 604 |
| 1D DEDICA | high vel. peak[a] | 45 | 54142 | > 538 |
| $< 0.3\,h_{70}^{-1}$ Mpc | E – clump | 18 | $54573 \pm 232$ | $944^{+303}_{-297}$ |
| $< 0.3\,h_{70}^{-1}$ Mpc | SW – clump | 8 | $53995 \pm 540$ | $1346^{+289}_{-102}$ |
| $< 0.3\,h_{70}^{-1}$ Mpc | NW – clump | 9 | $53487 \pm 407$ | $1105^{+182}_{-122}$ |
| 3D KMM | E – clump | 52 | $53908 \pm 162$ | $1155^{+119}_{-70}$ |
| 3D KMM | SW – clump | 17 | $53517 \pm 261$ | $1032^{+218}_{-106}$ |
| 3D KMM | NW – clump | 29 | $53312 \pm 166$ | $873^{+167}_{-91}$ |
| 3D KMM ($r' < 20$) | E – clump | 45 | $54077 \pm 170$ | $1128^{+123}_{-89}$ |
| 3D KMM ($r' < 20$) | SW – clump | 13 | $53079 \pm 244$ | $831^{+217}_{-110}$ |
| 3D KMM ($r' < 20$) | NW – clump | 30 | $53296 \pm 159$ | $852^{+165}_{-74}$ |

[a] For this galaxy clump we report the velocity peak instead on the mean velocity. The value of velocity dispersion should be considered a minor limit due to the abruptal truncation of the velocity distribution.